\documentclass[prl,twocolumn,aps,showpacs,superscriptaddress]{revtex4-1}
\usepackage{color}
\newcommand{\be}{\begin{equation}}
\newcommand{\bea}{\begin{eqnarray}}
\newcommand{\eea}{\end{eqnarray}}
\newcommand{\ee}{\end{equation}}
\usepackage{epsfig}
\usepackage{hyperref}
\usepackage[normalem]{ulem}
\begin{document}
\title{Strong Fulde-Ferrell Larkin-Ovchinnikov pairing fluctuations in polarized Fermi systems}
\author{Michele Pini}
\affiliation{School of Science and Technology, Physics Division, 
Universit\`a di  Camerino, 62032 Camerino (MC), Italy}
\author{Pierbiagio Pieri}
\affiliation{Dipartimento di Fisica e Astronomia, Universit\`a di Bologna, 40127 Bologna (BO), Italy}
\affiliation{INFN, Sezione di Bologna, 40127 Bologna (BO), Italy}
\author{G. Calvanese Strinati}
\affiliation{School of Science and Technology, Physics Division, 
Universit\`a di  Camerino, 62032 Camerino (MC), Italy}
\affiliation{INFN, Sezione di Perugia, 06123 Perugia (PG), Italy}
\affiliation{CNR-INO, Istituto Nazionale di Ottica, Sede di Firenze, 50125 Firenze (FI), Italy}
\date{\today}

\begin{abstract}
We calculate the pair susceptibility of an attractive spin-polarized Fermi gas in the normal phase, as a function of the pair momentum. 
Close to unitarity, we find a strong enhancement of Fulde-Ferrell-Larkin-Ovchinnikov (FFLO) pairing fluctuations over an extended region of the temperature-polarization phase diagram, 
which manifests itself as a pronounced peak in the pair-momentum distribution at a finite pair momentum. 
This peak should be amenable to experimental observation at achievable temperatures in a box-like trapping potential, as a fingerprint of FFLO pairing. 
Our calculations rest on a self-consistent $t$-matrix approach which, for the unitary balanced Fermi gas, has been validated against experimental data for several thermodynamic quantities.
 \end{abstract}
\maketitle

{\em Introduction $-$} 
One of the main motivations that some time ago~\cite{Zwierlein-2006,Partridge-2006} initiated a vibrant experimental and theoretical research activity on ultracold polarized attractive Fermi gases~\cite{Reviews-imbalanced},  was the search for exotic superfluid phases, like the FFLO phase predicted many years before by Fulde and Ferrell~\cite{Fulde-1964} and Larkin and Ovchinnikov~\cite{Larkin-1964} for superconductors in a strong magnetic field.  
This phase is characterized by pairing at finite center-of-mass momentum of the pairs, with spatially dependent gap parameter $\Delta({\bf r})= \Delta_{Q_0} e^{i{\bf Q}_0\cdot {\bf r}}$ for the Fulde-Ferrell (FF) and 
$\Delta({\bf r})= \Delta_{Q_0} \cos({\bf Q}_0\cdot {\bf r})$ for the Larkin-Ovchinnikov (LO) solutions. 
However, despite many efforts in ultracold gases, some (indirect) evidence for the FFLO phase was obtained so far in one dimension only~\cite{Liao-2010}.
In the mean time, substantial progress in the quest for the FFLO phase was made in condensed-matter systems, producing solid evidence of an FFLO phase in quasi-two-dimensional organic superconductors~\cite{Lortz-2007,Koutroulakis-2016,Wosnitza-2018}, 
as well as quite robust evidence in the iron-based multi-band superconductor KFe$_2$As$_2$~\cite{Cho-2017}. 
In addition, mounting evidence is emerging in cuprate superconductors \cite{Hamidian-2016,Edkins-2019,Du-2020,Agterberg-2020} about the presence of ``pair-density-wave" ordering (corresponding to an LO solution in the absence of a magnetic field),
which competes with $d$-wave superconductivity and possibly explains the pseudogap phase observed in these systems as associated with strong pair-density-wave fluctuations~\cite{Agterberg-2020}. 

In the present work, we consider the normal (i.e., not superfluid) phase of a polarized Fermi gas, and search for the presence of FFLO pair fluctuations by calculating the pair susceptibility vs pair momentum ${\bf Q}$.
Close to unitarity~\cite{Zwerger-2012}, we find that the pair susceptibility is considerably enhanced at a finite value $Q_{0}$ of $Q (=|{\bf Q}|)$ over an extended region of the temperature-polarization phase diagram, such that it should be possible to observe this strong tendency towards FFLO ordering at experimentally achievable temperatures (say, a few percent of the Fermi temperature) in the normal phase. 
To this end, we suggest that the projection technique introduced some time ago to detect pair condensation in a strongly interacting Fermi gas~\cite{Regal-2004,Zwierlein-2004} could as well be used here to measure a ``projected'' pair-momentum distribution,  
whereby the occurrence of a pronounced peak at the same $Q_{0}$ of the pair susceptibility would provide unambiguous evidence for strong FFLO pair fluctuations. 
We explicitly calculate this projected pair-momentum distribution, and conclude that its peak at a finite $Q_{0}$ should most readily be observed with a box-like trapping potential~\cite{Mukherjee-2017,Hueck-2018,Yan-2019,Shkredov-2021}.

 {\em Pair susceptibility $-$}
 We consider a system of spin-$\frac{1}{2}$ fermions of mass $m$ mutually interacting through a contact interaction, as described by the Hamiltonian (throughout, the reduced Planck constant $\hbar$ and the Boltzmann constant $k_B$ are set equal to unity): 
 \bea
H&=&\sum_{\sigma}\int \! d {\bf r} \, \hat{\psi}^{\dagger}_{\sigma}({\bf r}) \left( - \frac{\nabla^2}{2m} \right) \hat{\psi}_{\sigma}({\bf r}) \nonumber \\
&+& v_0\int \! d {\bf r} \, \hat{\psi}^{\dagger}_{\uparrow}({\bf r}) \hat{\psi}^{\dagger}_{\downarrow}({\bf r}) \hat{\psi}_{\downarrow}({\bf r}) \hat{\psi}_{\uparrow}({\bf r}),
\eea
 where $\hat{\psi}_{\sigma}({\bf r})$ is a field operator with spin projection $\sigma=(\uparrow,\downarrow)$ and $v_0$ the bare interaction strength 
 (with $v_0\to0^{-}$ when the contact interaction is regularized in terms of the two-fermion scattering length $a_{\rm F}$~\cite{Pieri-2000}).
 Our aim is to calculate the pair susceptibility $\chi_{\rm pair}({\bf Q})$, that describes the tendency of the normal Fermi gas  towards superfluid ordering with pair center-of-mass momentum ${\bf Q}$.
 This quantity can be obtained as  $\chi_{\rm pair}({\bf Q})={\cal D}({\bf Q},\Omega_{\nu}=0)$ \cite{Supp}, where 
\bea
{\cal D}({\bf Q},\Omega_{\nu})&=&\int_0^{1/T}\!\!\!\!d(\tau-\tau') e^{i\Omega_{\nu}(\tau-\tau')} \int \!\!d({\bf r}-{\bf r}') e^{-i {\bf Q}\cdot ({\bf r}-{\bf r}')}\nonumber\\
&\times& {\cal D}({\bf r}-{\bf r}',\tau-\tau')
\eea
is the Fourier transform of the response function \cite{Pistolesi-1996,Palestini-2014}
\be
{\cal D}({\bf r}-{\bf r}',\tau-\tau')=\langle T_{\tau}[\hat{\Delta}({\bf r}\tau)\hat{\Delta}^{\dagger}({\bf r}'\tau')]\rangle 
\ee
at the Matsubara frequency $\Omega_{\nu}=2\pi\nu T$ ($\nu$ integer). 
Here, $\langle  \cdots \rangle$ is a grand-canonical thermal average at temperature $T$, $\hat{\Delta}({\bf r}\tau)\equiv e^{\tau \hat{K}}\hat{\Delta}({\bf r})e^{-\tau \hat{K}}$ the gap operator at imaginary time $\tau$ with 
$\hat{\Delta}({\bf r})=v_0 \hat{\psi}_{\uparrow}({\bf r}) \hat{\psi}_{\downarrow}({\bf r})$,
and $\hat{K} = \hat{H} - \sum_{\sigma}\mu_{\sigma} \hat{N}_{\sigma}$ ($\mu_{\sigma}$  and $\hat{N}_{\sigma}$ being the chemical potential and the number operator for spin species $\sigma$, respectively).

{\em Self-consistent $t$-matrix approximation for a polarized Fermi gas $-$}
The simplest physically meaningful approximation for ${\cal D}({\bf Q},\Omega_{\nu})$ results by summing the series of ladder diagrams, yielding ${\cal D}({\bf Q},\Omega_{\nu})=\Gamma_0({\bf Q},\Omega_{\nu})$ \cite{Palestini-2014}.
Here, 
\bea
&&\Gamma_0({\bf Q},\Omega_{\nu})=-\left\{\frac{m}{4\pi a_{\rm F}} + \int\frac{d{\bf k}}{(2\pi)^3}\right.\nonumber\\
&&\left.\left[T\sum_n G^0_{\uparrow}({\bf k},\omega_n)G^0_{\downarrow}({\bf Q}-{\bf k},\Omega_{\nu}-\omega_n)-\frac{m}{k^2}\right]\right\}^{-1}
\label{gamma0}
\eea
is the bare pair propagator, where $G^0_{\sigma}({\bf k},\omega_n)=(i\omega_n-k^2/2m +\mu_{\sigma})^{-1}$ is the bare single-particle propagator with fermionic Matsubara frequency $\omega_n=\pi (2n+1) T$ ($n$ integer).
Within this approximation, the pair susceptibility $\chi_{\rm pair}({\bf Q})$ is then identified with $\Gamma_0({\bf Q},\Omega_{\nu}=0)$, such that  
\bea
\Gamma_0({\bf Q},\Omega_{\nu}=0)^{-1}=0 
\label{thouless}
\eea
would correspond to a diverging $\chi_{\rm pair}({\bf Q})$.
For ${\bf Q}=0$, the condition (\ref{thouless}) coincides with the standard BCS mean-field equation for the superfluid critical temperature $T_c$ (generalized here to the spin-imbalanced case), while for ${\bf Q}\neq 0$, it corresponds to the FFLO mean-field equation for $T_c$ obtained by setting $\Delta({\bf Q})\to 0$ in the corresponding gap equation (cf., e.g., Eq.~(51) of Ref.~\cite{Strinati-2018}).

 \begin{figure}[t]
\includegraphics[angle=0,width=8.3cm]{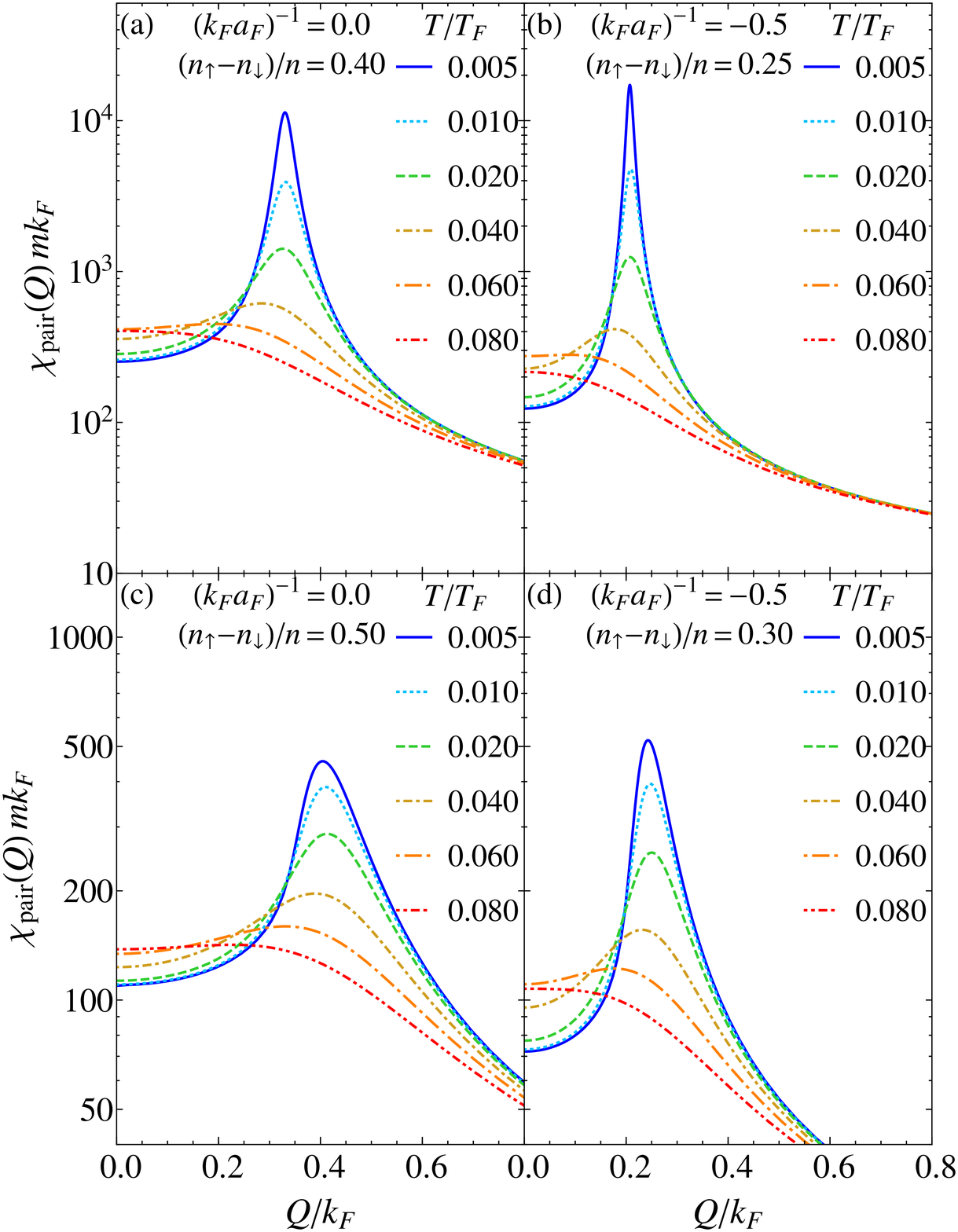}
\caption{Pair susceptibility $\chi_{\rm pair}(Q)$ (in units of $1/(m k_{\rm F})$) vs pair momentum $Q$ (in units of $k_{\rm F}$) at various temperatures. 
             Left panels refer to unitarity and polarizations $(n_{\uparrow}-n_{\downarrow})/n = 0.40$ (a) and 0.50 (c), while right panels refer to $(k_{\rm F} a_{\rm F})^{-1}=-0.5$ and polarizations 0.25 (b) and 0.30 (d).}
\label{Figure-1}
\end{figure}
 
 To obtain meaningful results for $T_c$ across the whole BCS-BEC crossover, the chemical potentials entering $\Gamma_0$ need to take into account the effects of pairing fluctuations in the normal phase~\cite{Nozieres-1985} 
 (in the balanced case, a single chemical potential $\mu_{\uparrow}=\mu_{\downarrow}=\mu$ survives). 
 This is achieved by inverting the density equations  $n_{\sigma}=\int \frac{d {\bf k}}{(2\pi)^3} T \sum_n G_{\sigma}({\bf k},\omega_n)e^{i\omega_n 0^+}$ in favor of the chemical potentials, where $G_{\sigma}$ is the single-particle propagator dressed by the $t$-matrix self-energy 
 $\Sigma_{\sigma}$, which is in turn constructed by convoluting the ladder series $\Gamma_0$ with a bare $G^0_{{\bar \sigma}}$~\cite{Perali-2002}
($\bar{\sigma}$ standing for the opposite of $\sigma$).

Comparison with experimental data or Quantum Monte Carlo (QMC) results for several thermodynamic quantities in the balanced case \cite{Sommer-2012,Zwerger-2016,Carcy-2019,Mukherjee-2019,Jensen-2020,Rammelmueller-2021}, as well as with recent QMC calculations for the polarized case \cite{Rammelmueller-2020},   shows, however, that a more reliable diagrammatic approximation is obtained by using a fully self-consistent $t$-matrix self-energy~\cite{Haussmann-1994,Haussmann-2007,Pini-2019}
\be
\Sigma_{\sigma}({\bf k},\omega_n)=-\int\!\!\frac{d{\bf Q}}{(2\pi)^3} T \sum_{\nu} \Gamma({\bf Q},\Omega_{\nu})G_{{\bar \sigma}}({\bf Q}-{\bf k},\Omega_{\nu}-\omega_n) \, ,
\label{self-consistent}
\ee
where $\Gamma({\bf Q},\Omega_{\nu})$ is defined by replacing $G^ 0_{\sigma} \rightarrow G_{\sigma}$ on the right-hand side of Eq.~(\ref{gamma0}).
By this approach, the pair susceptibility of interest is correspondingly given by 
\be
\chi_{\rm pair}({\bf Q})=\Gamma({\bf Q},\Omega_{\nu}=0) \, ,
\label{chi-gamma}
\ee
which retains two-particle diagrams consistent with the fully self-consistent $t$-matrix self-energy (\ref{self-consistent}) \cite{Supp}.
Our calculations of the pair susceptibility will be based on Eq.~(\ref{chi-gamma}) and on the self-consistent solution of Eq.~(\ref{self-consistent}).
 
 \begin{figure}[t]
\includegraphics[angle=0,width=8.8cm]{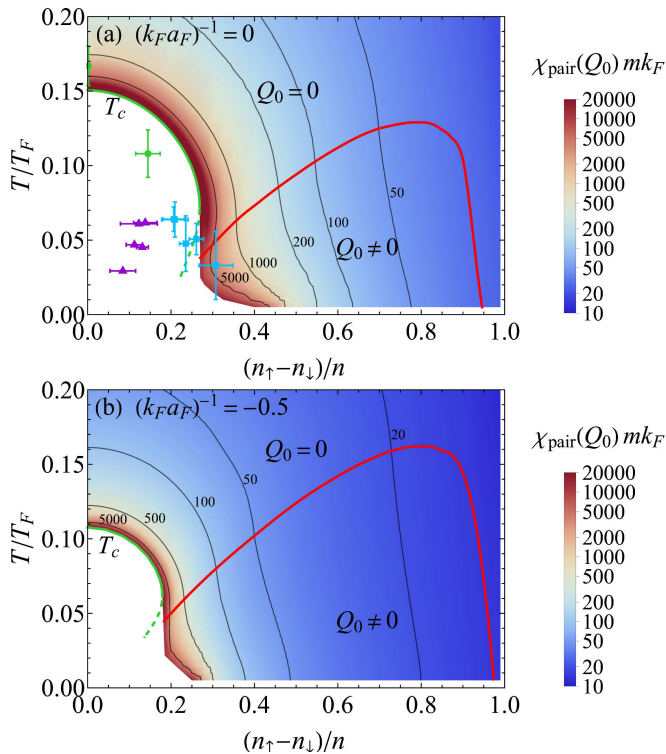}
\caption{Heat map for the peak value of the dimensionless pair susceptibility $\chi_{\rm pair}(Q) m k_{\rm F}$ in the temperature-polarization phase diagram, for couplings $(k_{\rm F} a_{\rm F})^{-1}=0$ (a) and -0.5 (b).   
             Contour lines for $\chi_{\rm pair} (Q_{0}) m k_{\rm F}$ are also shown. 
             The red curve delimits the region where the peak of $\chi_{\rm pair}(Q)$ occurs at $Q_0\neq 0$, while the green line corresponds to $\Gamma({\bf 0},0)^{-1}=0$.   
             Also reported in panel (a) are the experimental data~\cite{Shin-2008} for various phase transitions:  
             uniform normal - uniform superfluid (circles), uniform superfluid - phase separated (triangles),  and phase separated - uniform normal (squares).}
\label{Figure-2}
\end{figure}
 
{\em Numerical results $-$}
In the following, it will be useful to introduce the (effective) Fermi wave vector $k_{\rm F}\equiv(3\pi^2 n)^{1/3}$ defined in terms of the total density $n=n_{\uparrow} + n_{\downarrow}$.  
The dimensionless coupling $(k_{\rm F} a_{\rm F})^{-1}$ then drives the crossover from the BCS and BEC regimes, which correspond to $(k_{\rm F} a_{\rm F})^{-1}\lesssim -1$ and $(k_{\rm F} a_{\rm F})^{-1}\gtrsim +1$, respectively, 
while the crossover region in between spans across the unitarity limit $(k_{\rm F} a_{\rm F})^{-1}=0$. 
A dimensionless pair susceptibility is then defined as $\chi_{\rm pair}(Q)m k_{\rm F}$.
 
Figure~\ref{Figure-1} reports our results for $\chi_{\rm pair}(Q)m k_{\rm F}$ for two coupling values in the crossover region, namely, unitarity (left panels) and $(k_{\rm F} a_{\rm F})^{-1}=-0.5$ (right panels). 
For each coupling, two different polarizations are also considered. 
From these results one sees that the pair susceptibility gets strongly enhanced about a finite momentum $Q_0$ as the temperature is progressively lowered, thereby signaling the presence of strong FFLO fluctuations in the normal phase.  
This peak at $Q_0$ is rather well pronounced for temperatures $T\approx 0.05 T_{\rm F}$ where $T_{F}=k_{F}^{2}/(2m)$, which are well within the current experimental capabilities with ultracold Fermi gases.

To characterize location and strength of FFLO fluctuations in the temperature-polarization phase diagram, a heat map for the peak value of the dimensionless pair susceptibility $\chi_{\rm pair}(Q) m k_{\rm F}$ is presented in Fig.~\ref{Figure-2}, 
for the same couplings $(k_{\rm F} a_{\rm F})^{-1}=0$ (a) and -0.5 (b) considered in Fig.~\ref{Figure-1}. 
Remarkably, for both couplings the region of the phase diagram where the maximum of the pair susceptibility develops at a finite momentum $Q_0$ is quite sizable, reaching temperatures as high as about $0.15 T_{\rm F}$ and encompassing a wide range of polarization.  
In Fig.~\ref{Figure-2}, the strength of pairing fluctuations, as quantified by the peak heights from Fig.~\ref{Figure-1}, is indicated by the color code and contour lines.  
Note that, in the region where the peak of $\chi_{\rm pair}(Q)$ occurs at $Q_0=0$, the condition $\Gamma({\bf 0},0)^{-1}=0$ corresponding to a diverging $\chi_{\rm pair}(0)$ can be exactly satisfied (green line). 
This curve yields the transition temperature $T_c$ to a uniform polarized superfluid with pairing at $Q_0=0$ until it becomes reentrant.  
The reentrant part of the curve (green dashed line) is instead expected to be superseded by phase separation or by an FFLO superfluid phase~\cite{Frank-2018}.

\begin{figure}[t]
\includegraphics[angle=0,width=8cm]{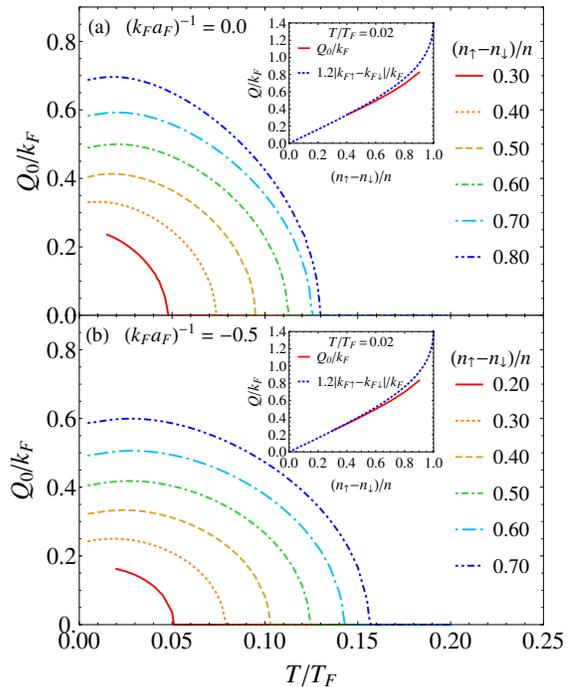}
\caption{Peak momentum $Q_0$ (in units of $k_{\rm F}$) occurring in the pair susceptibility $\chi_{\rm pair}(Q)$ as a function of temperature $T$ (in units of $T_{\rm F}$), 
             for the coupling values $(k_F a_F)^{-1}=0$ (a) and -0.5 (b) and a set polarizations $(n_{\uparrow}-n_{\downarrow})/n$.  
             For the same couplings, the insets compare the dependence of $Q_0$ on the polarization at the temperature $T=0.02 T_{\rm F}$ (full lines) against the function $1.2 |k_{{\rm F}\uparrow}-k_{{\rm F}\downarrow}|/k_{\rm F}$ (broken lines).}
\label{Figure-3}
 \end{figure}
 
When $Q_0\neq 0$, on the other hand, the feedback of a diverging $\chi_{\rm pair}(Q_0)$ on Eq.~(\ref{self-consistent}) would yield a diverging self-energy at finite temperatures for all frequencies and momenta 
(see Refs.~\cite{Shimahara-1998,Shimahara-1999,Ohashi-2002,Jakubczyk-2017,Wang-2018} and \cite{Strinati-2018} for a discussion of a similar phenomenon in related approaches). 
Accordingly, within the present approach the condition $\Gamma(Q_0,0)^{-1}=0$ (corresponding to a second-order phase transition to the FFLO phase) can be exactly satisfied only at $T=0$.
Recall, however, that, in the superfluid phase, FFLO fluctuations would turn FFLO ordering from long-range into algebraic~\cite{Radzihovsky-2011}, in such a way that determining the transition line by diagrammatic methods would be an extremely difficult task
(like for the Berezinskii-Kosterlitz-Thouless transition in a 2D superfluid Fermi gas). 
Nevertheless, this remark does not hinder our investigation, which focuses rather on the effects of the FFLO fluctuations in the normal phase than on the precise determination of the second-order transition line. 
In this respect, the experimental results for the unitary Fermi gas (symbols in Fig.~\ref{Figure-2}(a)) indicate that at low temperature the transition between normal and superfluid phases becomes actually of first order
(with a phase separation between a balanced superfluid and a normal polarized gas). 
Our calculations show that strong FFLO pairing fluctuations are expected to occur immediately at the right of this phase separation region found in the experiments.

Figure \ref{Figure-3} complements the results of Fig.~\ref{Figure-2}, by reporting for the same couplings the peak momentum $Q_0$ as a function of temperature for a set of polarizations.  
Here, $Q_0$ is seen to acquire a finite value in a continuous way when entering the FFLO fluctuation region and to further increase as the temperature decreases (except for a slight decrease at the lowest temperatures).
Recall in this context that the mean-field solution for the FF phase suggests that at low $T$ (where sharp Fermi surfaces develop) $Q_0$ should scale with the mismatch $k_{{\rm F}\uparrow}-k_{{\rm F}\downarrow}$ between the corresponding Fermi wave vectors. 
This expectation is confirmed by the comparison shown in the insets of Fig.~\ref{Figure-3}, between the polarization dependences of $Q_0$ (obtained at the low temperature $T=0.02 T_{\rm F}$) and 
the weak-coupling mean-field result $1.2(k_{{\rm F}\uparrow}-k_{{\rm F}\downarrow})$ for $Q_0$ at the $T=0$ superfluid-normal transition \cite{Takada-1969}.
 
\begin{figure}[t]
\includegraphics[angle=0,width=8cm]{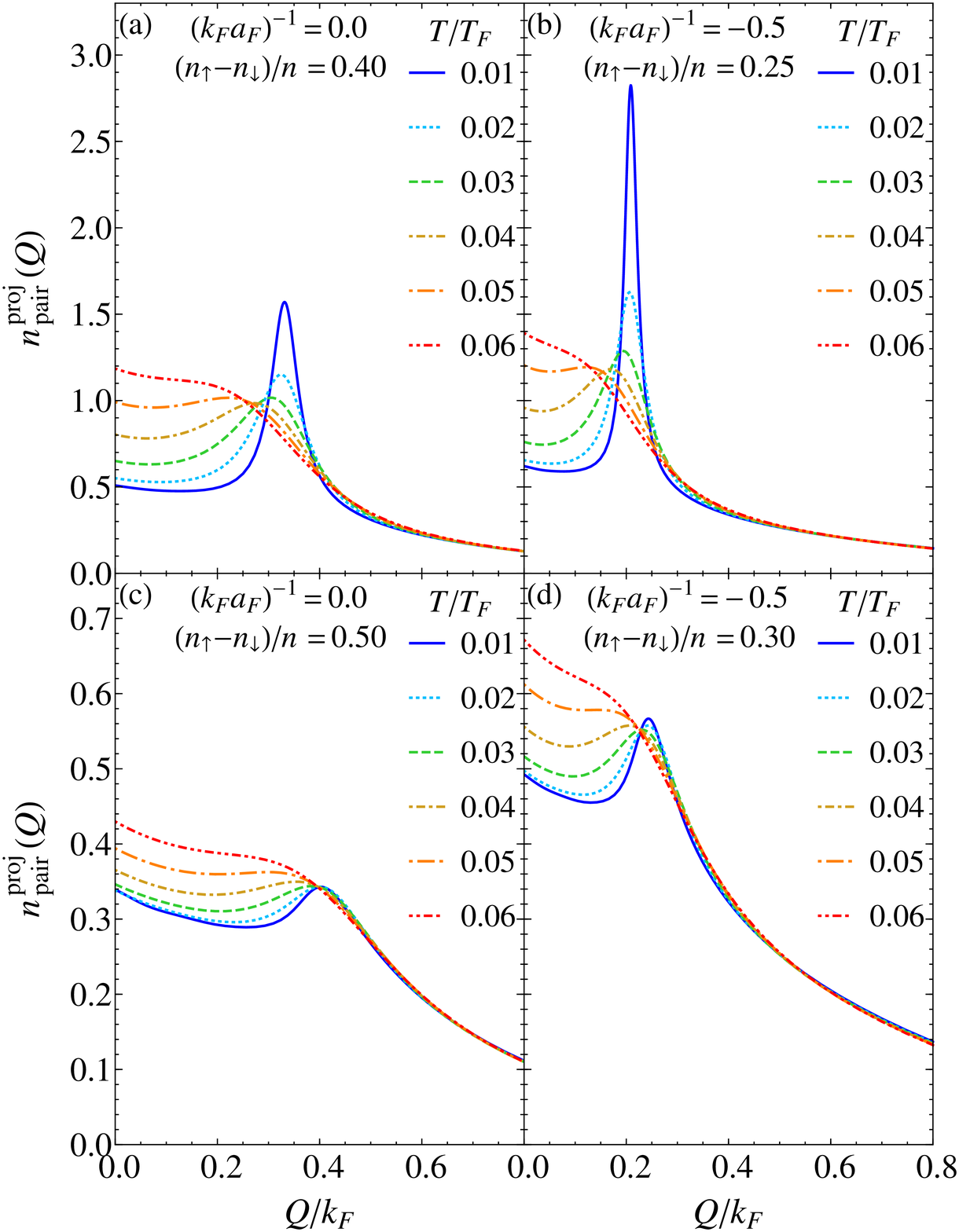}
\caption{Projected pair-momentum distribution $n^{\rm proj}_{\rm pair}(Q)$ as a function of the pair momentum $Q$ (in units of $k_{\rm F}$) for different temperatures. 
             The left panels show results at unitarity for polarizations $(n_{\uparrow}-n_{\downarrow})/n = 0.40$ (a) and 0.50 (c), while the right panels correspond to $(k_{\rm F} a_{\rm F})^{-1}=-0.5$ and polarizations 0.25 (b) and 0.30 (d).}
\label{Figure-4}
 \end{figure}

{\em Proposed experimental test -}  
One expects the pair susceptibility $\chi_{\rm pair}(Q)$ to hardly be measured in a direct way. 
A somewhat related quantity of more direct access to experiments with ultracold gases should be the ``projected'' pair-momentum distribution $n^{\rm proj}_{\rm pair}(Q)$, which is the momentum distribution of the molecules formed after a rapid sweep of the magnetic field to the BEC side of the crossover. 
Measurements of the projected pair-momentum distribution have already been successfully applied to detect condensation (or quasi-condensation) of fermionic pairs across the BCS-BEC crossover, 
both in three \cite{Regal-2004,Zwierlein-2004,Shkredov-2021,Dyke-2021} and two \cite{Ries-2015} dimensions, and were also proposed some time ago to detect FFLO superfluidity in trapped Fermi gases \cite{Yang-2005}. 
On physical grounds, we expect that, even in the normal phase, strong FFLO pairing fluctuations should result into a peak of $n^{\rm proj}_{\rm pair}(Q)$ at the same finite $Q_{0}$ found for $\chi_{\rm pair}(Q)$ \cite{footnote-lattice}. 
 
To confirm this expectation, we have extended to the normal phase the theoretical approach for the projected pair-momentum distribution introduced in Ref.~\cite{Perali-2005} for the superfluid phase, borrowing also from the formalism recently used in Ref.~\cite{Pini-2020} 
to address pair correlations in the normal phase of an attractive Fermi gas~\cite{Supp}.  
The results for $n^{\rm proj}_{\rm pair}(Q)$ are reported in Fig.~\ref{Figure-4} and show that, in the strong FFLO pairing-fluctuation region, a pronounced peak develops in $n^{\rm proj}_{\rm pair}(Q)$ at a finite momentum which matches the corresponding peak momentum $Q_0$ 
found in Fig.~\ref{Figure-1} for $\chi_{\rm pair}(Q)$.
The peak of $n^{\rm proj}_{\rm pair}(Q)$ is quite prominent at $T=0.01 T_{F}$ and remains clearly visible up to $T\approx 0.05 T_{\rm F}$, which is well within the range of temperatures currently attainable with ultracold gases \cite{footnote-pair-numbers}.
The results  of Fig.~\ref{Figure-4}   are for  a uniform system and thus  correspond to an idealization of a 
Fermi gas trapped in a box-like potential.
We have performed calculations \cite{Supp} also for a realistic box-like potential as the one used in Ref.~\cite{Mukherjee-2017}, confirming the  observability of the predicted peak even in this case.
  For a gas trapped in a harmonic potential,  on the other hand, we have verified that   it would be hard to detect a peak of $n^{\rm proj}_{\rm pair}(Q)$ at a finite $Q_0$, due to the smearing produced by the trap average \cite{Supp}.

{\em  Conclusions -}  
We have uncovered the presence of strong FFLO fluctuations in the normal phase of a polarized Fermi gas, which could experimentally be  observable even in a three-dimensional unitary Fermi gas of most interest for the current research in the field. 
These fluctuations are precursors of an FFLO superfluid phase, which is competing with the phase separation observed so far in experiments \cite{Zwierlein-2006,Shin-2008,Olsen-2015}. 
The outcome of this competition in the $T\to 0$ limit cannot be established from the experimental data, since phase separation could eventually either give the way to an FFLO superfluid phase or continue to suppress it entirely. 
Nevertheless, irrespective of which one of these two scenarios would actually take place, our investigation has shown that the effects of an underlying FFLO superfluid phase are clearly visible, and should accordingly be sought for, 
\emph{in the normal phase}.
 
 \begin{acknowledgments}
Financial support from the Italian MIUR under Project PRIN2017 (20172H2SC4) is acknowledged.
\end{acknowledgments}

\newpage
\section{Supplemental material}

\subsection{Pair susceptibility}
We consider a system of spin-$\frac{1}{2}$ fermions of mass $m$ mutually interacting through a contact interaction, as described by the Hamiltonian (with the reduced Planck constant $\hbar$ and the Boltzmann constant $k_B$ set equal to unity): 
 \bea
\hat{H}&=& \sum_{\sigma}\int d {\bf r} \, \hat{\psi}^{\dagger}_{\sigma}({\bf r}) \left( -\frac{\nabla^2}{2m} \right) \hat{\psi}_{\sigma}({\bf r}) \nonumber \\
&+& v_0\int d {\bf r} \, \hat{\psi}^{\dagger}_{\uparrow}({\bf r}) \hat{\psi}^{\dagger}_{\downarrow}({\bf r}) \hat{\psi}_{\downarrow}({\bf r}) \hat{\psi}_{\uparrow}({\bf r}) \, .
\label{H}
 \eea
Here, $\hat{\psi}_{\sigma}({\bf r})$ is a field operator with spin projection $\sigma=(\uparrow,\downarrow)$  and $v_0$ is the bare interaction strength ($v_0\to0^{-}$ when the contact interaction is regularized in terms of
two-fermion scattering length $a_{\rm F}$~\cite{Pieri-2000-S}).
 
To quantify the tendency of the normal Fermi gas towards superfluid ordering with pair center-of-mass momentum ${\bf Q}$, we consider an associated pair susceptibility $\chi_{\rm pair}({\bf Q})$ defined in the following way.
 We add to the Hamiltonian (\ref{H}) the symmetry-breaking term 
\be
\hat{H}_{\rm ext}=   -\int d{\bf r} \, \eta({\bf r}) \, \hat{\varphi}({\bf r}) \, ,
\label{Hex2}
\ee
where 
\be
\hat{\varphi}({\bf r}) = \left( \frac{ \hat{\Delta}({\bf r}) + \hat{\Delta}^{\dagger}({\bf r}) }{\sqrt{2}} \right)
\label{phi}
\ee
and $\eta({\bf r})$ is a classical (real) field coupled to the gap operator $\hat{\Delta}({\bf r})=v_0 \psi_{\uparrow}({\bf r}) \psi_{\downarrow}({\bf r})$ and its hermitian conjugate $\hat{\Delta}^{\dagger}({\bf r})$.
Within linear-response theory, the pair susceptibility $\chi_{\rm pair}({\bf Q})$ is then obtained as the Fourier transform 
\be
\chi_{\rm pair}({\bf Q}) = \int d({\bf r}-{\bf r}') e^{-i {\bf Q}\cdot ({\bf r}-{\bf r}')}\chi_{\rm pair}({\bf r}-{\bf r}')
\ee
of the local functional derivative
\be
\chi_{\rm pair}({\bf r}-{\bf r}') = \left. \frac{\delta \langle \hat{\varphi}({\bf r})\rangle_{\eta} }{\delta \eta({\bf r}')}\right|_{\eta=0} \, .
\label{chi_gen}
\ee
Here, $ \langle \hat{\mathcal{A}} \rangle_{\eta} = \frac{ \mathrm{Tr} \{ e^{-\beta \hat{K}} \hat{\mathcal{A}} \}}{\mathrm{Tr} \{ e^{-\beta \hat{K}} \}}$
stands for the grand-canonical thermal average of a generic operator $\hat{\mathcal{A}}$, with the grand-canonical Hamiltonian $\hat{K} = \hat{H} + \hat{H}_{\rm ext} - \sum_{\sigma}\mu_{\sigma} \hat{N}_{\sigma}$ containing the symmetry-breaking term (\ref{Hex2}), and $\beta=1/T$ is the inverse temperature.  

When $\eta({\bf r})$ kept finite (albeit small), the expression (\ref{chi_gen}) implies that a weak probing field of the form $\eta({\bf r}) = \eta \cos({\bf Q}_0\cdot {\bf r})$ induces in the normal phase a gap parameter $\Delta_{Q_0}\!(\mathbf{r})$ proportional to $\chi_{\rm pair}(Q_0) \, \eta \cos({\bf Q}_0\cdot {\bf r})$, thus signaling that $\chi_{\rm pair}(Q)$ quantifies the tendency towards FFLO pairing (with rotational invariance implying that $\chi_{\rm pair}({\bf Q})$ is a function of $Q = |{\bf Q}|$). 
Accordingly, when $\chi_{\rm pair}(Q)$ is found to diverge at a finite momentum $Q_0$, a continuous transition develops from the normal phase to an FFLO phase with pair momentum $Q_0$.
(Recall in this respect that when the gap parameter $\Delta_{Q_0}\!(\mathbf{r}) \to 0$ from the superfluid phase with $\eta = 0$, as expected for a continuous phase transition, the FF and LO solutions  become degenerate, such that the transition point is the same for the two phases \cite{Larkin-1964-S}.)
More generally, evidence that $\chi_{\rm pair}(Q)$ becomes strongly peaked about a finite value $Q_0$ of $Q$ can be considered as indicating the presence of strong FFLO pairing fluctuations in the normal phase.

The pair susceptibility (\ref{chi_gen}) can be related to an appropriate temperature response function.
This is achieved by calculating the functional derivative in Eq.~(\ref{chi_gen}) as 
\be
\frac{\delta \langle \hat{\varphi}({\bf r})\rangle_{\eta} }{\delta \eta({\bf r}')} = \frac{ \mathrm{Tr} \{  \frac{\delta e^{- \beta \hat{K}} }{\delta \eta({\bf r}')} \hat{\varphi}({\bf r}) \} }{ \mathrm{Tr} \{ e^{- \beta \hat{K}} \} }  -
\frac{ \mathrm{Tr} \{  e^{- \beta \hat{K}} \hat{\varphi}({\bf r}) \} }{ \mathrm{Tr} \{ e^{- \beta \hat{K}} \} }
\frac{ \mathrm{Tr} \{  \frac{\delta e^{- \beta \hat{K}} }{\delta \eta({\bf r}')} \} }{ \mathrm{Tr} \{ e^{- \beta \hat{K}} \} }
\label{functional-derivative}
\ee
with $\eta$ still kept finite, and using the following operator identity
\be
e^{( \hat{\mathcal{A}} + \delta \hat{\mathcal{A}} ) s} = e^{ \hat{\mathcal{A}} s} \, \left[  1 + \int_{0}^{s} \!\!\! d s' \, e^{ - \hat{\mathcal{A}} s'} \delta  \hat{\mathcal{A}} \, e^{ - \hat{\mathcal{A}} s'} + \cdots \right] 
\label{operator-identity}
\ee
to linear order in $\delta \hat{\mathcal{A}}$.
In our case, $\hat{\mathcal{A}} \leftrightarrow - \hat{K}$ and $s \leftrightarrow \beta$, such that in Eq.~(\ref{functional-derivative})
\be
\frac{\delta e^{- \beta \hat{K}} }{\delta \eta({\bf r}')} = e^{- \beta \hat{K}} \, \int_{0}^{\beta} \!\!\! d\tau' \, e^{\tau' \hat{K}} \hat{\varphi}(\mathbf{r}') e^{- \tau' \hat{K}} \, .
\label{functional-identity-partial}
\ee
With the definition $\hat{\varphi}(\mathbf{r},\tau) = e^{\tau \hat{K}} \hat{\varphi}(\mathbf{r}) e^{- \tau \hat{K}}$, Eq.~(\ref{functional-derivative}) becomes eventually:
\be
\frac{\delta \langle \hat{\varphi}({\bf r})\rangle_{\eta} }{\delta \eta({\bf r}')} = \int_{0}^{\beta} \!\!\! d\tau' \, \langle \hat{\varphi}(\mathbf{r}',\tau') \hat{\varphi}(\mathbf{r}) \rangle_{\eta} 
                                                                                                                   - \beta \, \langle \hat{\varphi}(\mathbf{r}) \rangle_{\eta} \, \langle \hat{\varphi}(\mathbf{r}') \rangle_{\eta} \, .
\label{functional-identity-final}
\ee
In the normal phase of interest $\langle \hat{\varphi}(\mathbf{r}) \rangle_{\eta \rightarrow 0} = 0$, such that within linear response the local pair susceptibility (\ref{chi_gen}) can be expressed in terms of the temperature response function
\be
{\cal D}_{\varphi}({\bf r} \tau, {\bf r}' \tau') = \langle T_{\tau} [ \hat{\varphi}({\bf r}\tau) \, \hat{\varphi}({\bf r}'\tau')] \rangle_{\eta=0} 
\label{temperature-response-function}
\ee
where $T_{\tau}$ is the imaginary-time operator \cite{FW-1971-S}.
One obtains:
\bea
& & \chi_{\rm pair}({\bf r}-{\bf r}') = \!\! \int_{0}^{\beta} \!\!\!\! d(\tau - \tau') \, {\cal D}_{\varphi}({\bf r} - {\bf r}', \tau - \tau') 
\label{pair-susceptibility-final} \\
& = & \!\! \int_{0}^{\beta} \!\!\!\! d(\tau - \tau') \, {\cal D}({\bf r} - {\bf r}', \tau - \tau') = {\cal D}({\bf r} - {\bf r}',\Omega_{\nu} = 0)
\nonumber
\eea
where ${\cal D}({\bf r} - {\bf r}', \tau - \tau')$ is the function defined in Eq.~(3) of the main text and $\Omega_{\nu} = 2 \pi \nu / \beta$ ($\nu$ integer) is a bosonic Matsubara frequency \cite{FW-1971-S}.
To get Eq.~(\ref{pair-susceptibility-final}), homogeneity and isotropy in space and homogeneity in imaginary time have been exploited.
In Fourier space, one further obtains that $\chi_{\rm pair}({\bf Q}) = {\cal D}({\bf Q},\Omega_{\nu} = 0)$.

\subsection{Connection with many-body diagrammatic theory}

There remains to implement the calculation of ${\cal D}({\bf Q},\Omega_{\nu} = 0)$ in diagrammatic terms.
In this context, the simplest physically meaningful approximation results by summing the series of ladder diagrams in the particle-particle channel, where all rungs contain bare single-particle propagators.
Through a careful handling of the $v_{0} \rightarrow 0$ limit (where the interaction strength $v_{0}$ enters the definition of the gap operator $\hat{\Delta}(\mathbf{r})$), one can show
that ${\cal D}({\bf Q},\Omega_{\nu}) = \Gamma_{0}({\bf Q},\Omega_{\nu})$ with the bare pair propagator $\Gamma_{0}$ given by Eq.~(4) of the main text (cf. also Ref.~\cite{Palestini-2014-S}).

One knows, however, that an improved description of thermodynamic properties of a Fermi gas with attractive interaction results (at least in the balanced case) within the fully self-consistent $t$-matrix approximation \cite{Pini-2019-S}, 
where the bare $\Gamma_{0}$ is replaced in the single-particle self-energy $\Sigma$ by the dressed $\Gamma$ which contains the fully self-consistent single-particle propagators $G$ in the place of the the bare $G_{0}$.
It is then natural to replace $\Gamma_{0}$ with $\Gamma$ also in the expression of $\chi_{\rm pair}({\bf Q})$, as it was done in Eq.~(7) of the main text.

In this context, the question naturally arises about the need to introduce additional diagrammatic contributions to the temperature response function (\ref{temperature-response-function}) once the single-particle self-energy $\Sigma$ 
is taken within the fully self-consistent $t$-matrix approximation.
Specifically, we are referring to contributions with the topology of the Aslamazov-Larkin (AL) and Maki-Thompson (MT) diagrams, which, at the level of the fully self-consistent $t$-matrix approximation for the single-particle self-energy
here adopted, would influence the (two-particle) response functions through the particle-hole channel.
For the temperature response function (\ref{temperature-response-function}) of interest here, this would be the case if it were calculated in the superfluid phase below the critical temperature $T_{c}$, where there is no clear distinction
between particle-hole and particle-particle channels due to the particle-hole mixing characteristic of the BCS (pairing) theory.
However, since we are limiting ourselves to the normal phase above $T_{c}$, we can readily adapt to the present context the argument described in Appendix A of Ref.~\cite{Pini-2020-S}, and show that above $T_{c}$ the AL and MT contributions cannot affect the temperature response function (\ref{temperature-response-function}) due to its explicit spin structure and its ultimate particle-particle character.

\subsection{Projected pair-momentum distribution}

\subsubsection{Formalism and calculations for a homogeneous system}

The projected pair-momentum distribution can be obtained as follows in terms of the composite-boson propagator $G_{\rm B}(\mathbf{Q},\Omega_{\nu})$ \cite{Andrenacci-2003-S,Perali-2005-S}:
\begin{equation}
n^{\rm proj}_{\rm pair}(\mathbf{Q})=- T \sum_{\nu} G_{\rm B}(\mathbf{Q},\Omega_{\nu})e^{i\Omega_{\nu}0^+}.
\label{n_proj_pair}
\end{equation}
Within the self-consistent $t$-matrix approach, $G_{\rm B}(\mathbf{Q},\Omega_{\nu})$ is, in turn, given by \cite{Pini-2020-S}
\begin{equation}
G_{B}(\mathbf{Q},\Omega_{\nu})  =  - \mathcal{F}_{2}(\mathbf{Q},\Omega_{\nu}) - \mathcal{F}_{1}(\mathbf{Q},\Omega_{\nu})^{2} \, \Gamma(\mathbf{Q},\Omega_{\nu}) \, ,
\label{G_B-FT-sc}
\end{equation}
where $\mathcal{F}_{j}(\mathbf{Q},\Omega_{\nu})$  ($j=1,2$) are ``form factors'' defined as
\begin{eqnarray}
\mathcal{F}_{j}(\mathbf{Q},\Omega_{\nu}) & = & \int \! \frac{d{\mathbf p}}{(2\pi)^{3}} \, \phi_{\rm proj}({\mathbf p}+{\mathbf Q}/2)^{j}   
\label{def-F-j} \\
& \times & T \sum_{n} G_{\uparrow}(\mathbf{p}+\mathbf{Q},\omega_{n}+\Omega_{\nu}) G_{\downarrow}(-\mathbf{p},-\omega_{n}) .    
\nonumber
\end{eqnarray}
Here, $\phi_{\rm proj}(\mathbf{p})$ is the molecular wave function onto which the initial correlated pairs are projected during a magnetic sweep, while
$G_{\sigma}(\mathbf{p},\omega_{n})$ and $\Gamma(\mathbf{Q},\Omega_\nu)$ are the self-consistent single-particle Green's functions and the particle-particle (ladder) propagator defined in the main text.

The analysis of Ref.~\cite{Perali-2005-S} for projection experiments established that projection onto molecules occurs at a coupling  $(k_F a_F)^{-1}_{\rm proj}$ on the BEC side of the crossover, in order to optimize the overlap between the initial pair correlations and the molecular wave function.
 The specific value of the projection coupling $(k_F a_F)^{-1}_{\rm proj}$ depends on the experimental conditions of the magnetic sweep, and it was estimated to be generally in the range $0.5 \lesssim (k_F a_F)^{-1}_{\rm proj} \lesssim 1.5$
\cite{Perali-2005-S}.
Accordingly, for the calculation of $n^{\rm proj}_{\rm pair}(\mathbf{Q})$ shown in Fig.~4 of the main text we have considered a projection coupling $(k_F a_F)_{\rm proj}^{-1}=1$ in the middle of the above range. 
In addition, following again the procedure of Ref.~\cite{Perali-2005-S}, we have taken the molecular wave function  $\phi_{\rm proj}(\mathbf{p})$ as the normalized two-body bound-state wave function in vacuum at the projection coupling 
$(k_F a_F)^{-1}_{\rm proj}$:
\begin{equation}
\phi_{\rm proj}({\mathbf p})  =   k_{F}^{-3/2} \frac{\sqrt{8 \pi (k_{F} a_{F})^{-1}_{\rm proj}} }{({\mathbf p}/k_{F})^{2}  +  (k_{F} a_{F})_{\rm proj}^{-2}}  \, .
\label{psi-k-BEC}
\end{equation}

\begin{figure}[t]
\includegraphics[angle=0,width=8cm]{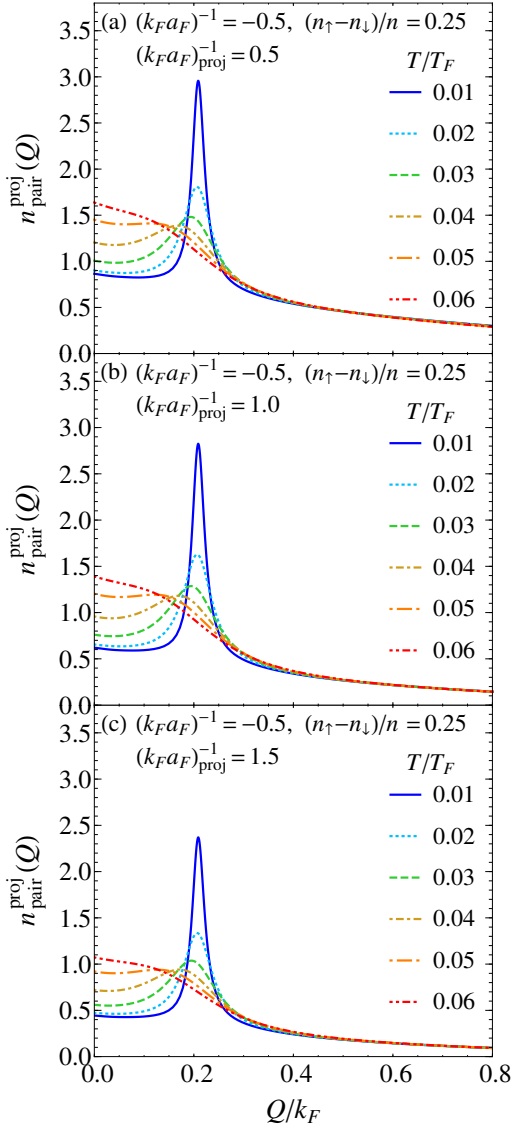}
\caption{Projected pair-momentum distribution $n_{\rm proj}^{\rm pair} (Q)$ vs pair momentum $Q$ (in units of $k_{\rm F}$) at different temperatures (in units of $T_{\rm F}$), 
              for $(k_{\rm F} a_{\rm F})^{-1}=-0.5$ and $(n_{\uparrow}-n_{\downarrow})/n = 0.25$. 
              The panels correspond to different values of the projection coupling $(k_F a_F)^{-1}_{\rm proj}\!: 0.5$ (a), 1.0 (b), and 1.5 (c).}
\label{Fig-S1}
\end{figure}

We have further verified that $n^{\rm proj}_{\rm pair}(\mathbf{Q})$ depends only weakly on the projection coupling  $(k_F a_F)^{-1}_{\rm proj}$. 
This is evident by comparing the results for $n^{\rm proj}_{\rm pair}(\mathbf{Q})$ shown in the three panels in Figure \ref{Fig-S1}, which correspond to different values of the projection coupling spanning the whole above range.

\subsubsection{Calculations for a harmonic trapping potential}

We next consider a trapped Fermi gas in a harmonic potential, which can be described by a local-density approximation. 
Accordingly, wherever they appear, we replace the chemical potentials $\mu_\sigma$ of the two $\sigma=(\uparrow,\downarrow)$ components by the local chemical potentials $\mu_\sigma (\mathbf{r})=\mu_{0 \sigma}-V(\mathbf{r})$, 
where $V(\mathbf{r})=m \omega_0^2 \mathbf{r}^2/2$ is the harmonic trapping potential with frequency $\omega_0$.  
By this replacement, the single-particle propagators $G_\sigma(\mathbf{k},\omega_n; \mathbf{r})$ and the composite-boson propagator $G_B(\mathbf{Q},\Omega_n; \mathbf{r})$ become \emph{local\/} functions of the position $\mathbf{r}$ 
in the trap through the local chemical potentials $\mu_\sigma(\mathbf{r})$.

For given particle numbers $N_{\sigma}$, the thermodynamic chemical potentials $\mu_{0 \sigma}$ (which correspond to the local chemical potentials at the trap center) are then obtained by inverting the number equations 
\begin{equation}
N_\sigma = \int \!\!d\mathbf{r} \int \!\!\frac{d\mathbf{k}}{(2\pi)^3} \, T \sum_n G_\sigma(\mathbf{k},\omega_n; \mathbf{r}) e^{i \omega_n 0^+} \, ;
\end{equation}
while the total projected pair-momentum distribution $N^{\rm proj}_{\rm pair}(\mathbf{Q})$ is obtained in the trap by summing the local projected pair-momentum distribution $n^{\rm proj}_{\rm pair}(\mathbf{Q}; \mathbf{r})$  over $\mathbf{r}$
 \begin{equation}
N^{\rm proj}_{\rm pair}(\mathbf{Q}) =   \int \!\!d \mathbf{r} \, n^{\rm proj}_{\rm pair}(\mathbf{Q}; \mathbf{r}) 
\label{n_proj_pair_trap_tot}
\end{equation}
where
\begin{equation}
n^{\rm proj}_{\rm pair}(\mathbf{Q}; \mathbf{r}) =  - \, T \sum_{\nu} G_{\rm B}(\mathbf{Q},\Omega_{\nu}; \mathbf{r})e^{i\Omega_{\nu}0^+} \, .
\label{n_proj_pair_local}
\end{equation}
To obtain an intensive quantity, one can then divide the total projected pair-momentum distribution of Eq.~(\ref{n_proj_pair_trap_tot}) by the volume $N/(k^t_F)^3$, where $k^t_F=\sqrt{2 m E^{t}_{F}}$ is the (effective) trap wave vector and $E^{t}_{F} = \omega_{0} (3N)^{1/3}$ is the (effective) trap Fermi energy, and obtain the projected pair-momentum distribution in the harmonic trap:
\begin{equation}
n_{\rm proj}^{{\rm pair}, h} (\mathbf{Q}) = \frac{(k^t_F)^3}{N} N^{\rm proj}_{\rm pair}(\mathbf{Q}) \, .
\end{equation}

\begin{figure}[t]
\includegraphics[angle=0,width=8cm]{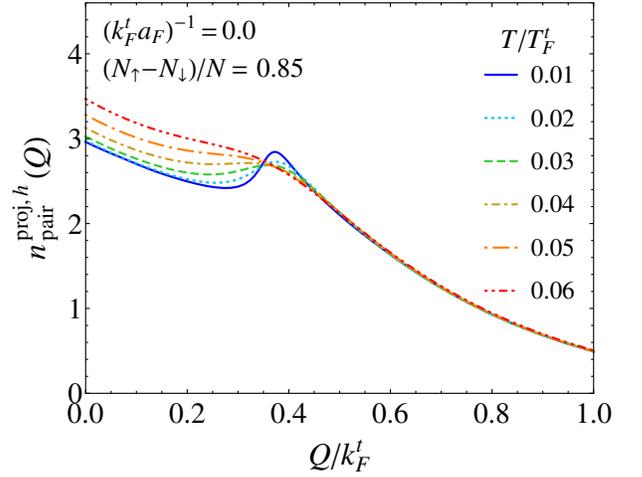}
\caption{Projected pair-momentum distribution in a harmonic trap $n_{\rm proj}^{{\rm pair}, h} (Q)$ vs pair momentum $Q$ (in units of $k_{\rm F}^t$) at unitarity and different temperatures, for a global
             polarization $(N_{\uparrow}-N_{\downarrow})/N = 0.85$ (which corresponds to a local polarization $(n_{\uparrow}-n_{\downarrow})/n \simeq 0.40$ at the trap center).}
\label{Fig-S2}
\end{figure}

Figure~\ref{Fig-S2} shows the results of $n_{\rm proj}^{{\rm pair}, h} (\mathbf{Q})$ obtained for a harmonic trap at unitarity and different temperatures, with global  polarization $(N_{\uparrow}-N_{\downarrow})/N=0.85$
where $N=N_{\uparrow}+N_{\downarrow}$.  
This value corresponds to a local polarization $(n_{\uparrow}-n_{\downarrow})/n \simeq 0.40$ at the trap center, which matches the value of the polarization considered in  Fig.~$4$(a) of the main text.
Here, the coupling parameter $(k^{t}_{F} a_{F})^{-1}$ and the pair momentum $Q$ are expressed in terms of the (effective) trap Fermi wave vector $k^{t}_{F}$, and the temperature $T$ is expressed in terms of the (effective) trap 
Fermi temperature $T^{t}_F=E^{t}_F$.

By comparing the results of Fig.~\ref{Fig-S2} for the trapped system with the corresponding results shown Fig.~$4$(a) in the main text for the  uniform \color{black}system, one is led to conclude that the effect of the harmonic trap is to strongly smear 
the peak at finite $Q$ which was clearly visible for the  uniform \color{black}system, making it hard to be detected experimentally. 
This smearing can be justified by extending to the trap the results reported in the inset of Fig.~3(b) of the main text, whereby the local value of the pair momentum $Q_0(\mathbf{r})$ at which $n^{\rm proj}_{\rm pair}(\mathbf{Q}; \mathbf{r})$ is peaked is bound to scale with $|k_{F \uparrow}(\mathbf{r})-k_{F \downarrow}(\mathbf{r})|$, which in turn varies along the trap since both local densities and polarization change with $\mathbf{r}$.
The net effect is that Eq.~(\ref{n_proj_pair_trap_tot}) effectively averages over locally projected pair-momentum distributions with different peak momenta $Q_0(\mathbf{r})$, thus smearing out the resulting peak in the projected pair-momentum distribution 
$n_{\rm proj}^{{\rm pair}, h} (\mathbf{Q})$ for the harmonic trap with respect to the corresponding peak that would be present for the  uniform \color{black}case.

\subsubsection{Calculations for a realistic box-like trapping potential}

It is relevant to calculate the projected pair-momentum distribution within a local-density approximation for a realistic box-like trapping potential. 
To this end, we consider the same kind of cylinder-shaped potential utilized in Ref.~\cite{Mukherjee-2017-S}. 
The trapping potential in cylindrical coordinates is given by $V(\rho,z)/E_F(0)=(\rho/R)^{p}  + \alpha_z (z/L)^2$ (in units of the (effective) Fermi energy $E_F(0)=[3 \pi^2 n(0)]^{2/3}/(2m)$, where $n(0)$ is the total density at the trap center).
Here, $V(\rho,z)$ is the sum of a radial power-law potential ($\sim \rho^{\,p}$), which describes the confinement of the ring beam, and of a weak axial harmonic potential (with $\alpha_z \ll 1$), which is needed to obtain the momentum distributions of atoms and pairs in a quarter period time-of-flight expansion \cite{Mukherjee-2017-S}. 
A hard-wall potential at $z=\pm L$ is also included to describe the light sheets acting as the end caps for the axial trapping. 
In Ref.~\cite{Mukherjee-2017-S} it was estimated that $p \simeq 16$ for the power-law exponent and $\alpha_z \le 0.05$ for the axial harmonic confinement parameter, while $L \simeq R$. 

\begin{figure}[t]
\includegraphics[angle=0,width=8cm]{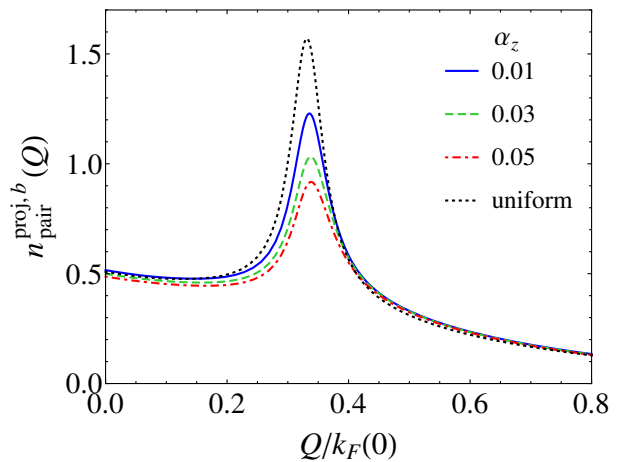}
\caption{The projected pair-momentum distribution $n^{{\rm proj}, b}_{\rm pair}(Q)$ obtained with a realistic box-like trap is shown as a function of the pair momentum $Q$ (in units of $k_{F}(0)=[3\pi^2 n(0)]^{1/3}$), 
             at unitarity and for $T/T_F(0)=0.01$.
             The calculations are done for various values of the axial harmonic confinement parameter $\alpha_z$, with local polarization $(n_{\uparrow}-n_{\downarrow})/n = 0.40$ 
             at the trap center (which corresponds to a global polarization $(N_{\uparrow}-N_{\downarrow})/N \simeq 0.47$). 
             The corresponding projected pair-momentum distribution for the uniform system is also shown for comparison (black dotted line).}
\label{Fig-S3}
\end{figure}

To perform a direct comparison with the uniform case, the (intensive) projected pair-momentum distribution for the box-like trap is obtained as
\begin{equation}
n^{{\rm proj}, b}_{\rm pair}(\mathbf{Q})= \frac{1}{V_{0}}  \int \!\!d \mathbf{r} \, n^{\rm proj}_{\rm pair}(\mathbf{Q}; \mathbf{r}) \, ,
\label{n_proj_pair_boxtrap}
\end{equation}
where the local $n^{\rm proj}_{\rm pair}(\mathbf{Q}; \mathbf{r})$ is defined like in Eq.~(\ref{n_proj_pair_local}) and $V_{0}=N/n(0)$ is the volume that would be occupied by the $N=N_\uparrow+N_\downarrow$ particles for a uniform system with density equal to the density $n(0)$ at the trap center. 
This is because, for a completely uniform trap such that $n^{\rm proj}_{\rm pair}(\mathbf{Q}; \mathbf{r})=n^{{\rm proj}}_{\rm pair}(\mathbf{Q})$ independent of $\mathbf{r}$, the integration over $\mathbf{r}$ in Eq.~(\ref{n_proj_pair_boxtrap}) would cancel the volume $V_0$ yielding $n^{{\rm proj}, b}_{\rm pair}(\mathbf{Q}) = n^{\rm proj}_{\rm pair}(\mathbf{Q})$.

Figure \ref{Fig-S3} shows the results for the projected pair-momentum distribution in a box-like trap with $p=16$ and $L=R$, corresponding to different values of the axial harmonic confinement parameter $\alpha_z \le 0.05$ like in Ref.~\cite{Mukherjee-2017-S}. 
For comparison, the corresponding result for the uniform system is also shown (cf. Fig.~4(a) of the main text). 
The calculations are performed by taking the local temperature and polarization at the trap center to coincide with those of the uniform system. 
From this comparison one concludes that the peak originating at finite momentum in the projected pair -momentum distribution for a uniform system still appears to be quite prominent even when the effects of a realistic box-like trapping potential are taken into account (and this is especially true when the axial harmonic trapping potential is kept sufficiently weak like in the experiment of Ref.~\cite{Mukherjee-2017-S}).
The favorable outcome of this specific test considerably reinforces our expectation as discussed in the main text, that the presence of strong FFLO fluctuations in the normal phase of a polarized Fermi gas can be uncovered by measurements of the projected pair-momentum distribution under realistic experimental conditions.


\begin{thebibliography}{99}
\bibitem{footnote-pair-numbers}  As a further information useful for future experiments, we mention that the expected fraction of pairs over the number of minority fermions is quite large, being e.g. about 0.55 for panel 4(a) and about 0.45 for panel 4(b).
\end{thebibliography}
\end{document}